\documentclass[%
aip,
reprint,
superscriptaddress,
ymb,
prb,
twocolumn,
titlepage,
floatfix,
showpacs,
]{revtex4-2}

\usepackage{graphicx}
\usepackage{hyperref}
\usepackage{amssymb}
\usepackage{amsmath}
\usepackage{textcomp}
\usepackage{xcolor}
\allowdisplaybreaks

\newcommand*{\bracite}[1]{%
  \textsuperscript{[}\cite{#1}\textsuperscript{]}}

\definecolor{linkblue}{RGB}{49,49,148}
\hypersetup{linkcolor  = linkblue,citecolor  = linkblue,urlcolor   = linkblue,colorlinks = true,}

\makeatletter
\renewcommand*{\eqref}[1]{%
  \hyperref[{#1}]{\textup{\tagform@{\ref*{#1}}}}%
}
\makeatother

\begin{document}

\title{Accelerating the laser-induced phase transition in nanostructured FeRh via plasmonic absorption}

\author{M.~Mattern}
\affiliation{Institut f\"ur Physik und Astronomie, Universit\"at Potsdam, 14476 Potsdam, Germany}
\author{J.-E.~Pudell}
\affiliation{Institut f\"ur Physik und Astronomie, Universit\"at Potsdam, 14476 Potsdam, Germany}
\affiliation{Helmholtz-Zentrum Berlin f\"ur Materialien und Energie GmbH, Wilhelm-Conrad-R\"ontgen Campus, BESSY II, 12489 Berlin, Germany}
\affiliation{European XFEL, 22869 Schenefeld, Germany}
\author{J. A.~Arregi}
\email{arregi@vutbr.cz}
\affiliation{CEITEC BUT, Brno University of Technology, 61200 Brno, Czech Republic}
\author{J.~Zl\'{a}mal}
\affiliation{CEITEC BUT, Brno University of Technology, 61200 Brno, Czech Republic}
\affiliation{Institute of Physical Engineering, Brno University of Technology, 61669 Brno, Czech Republic}
\author{R.~Kalousek}
\affiliation{CEITEC BUT, Brno University of Technology, 61200 Brno, Czech Republic}
\affiliation{Institute of Physical Engineering, Brno University of Technology, 61669 Brno, Czech Republic}
\author{V.~Uhl\'{i}\v{r}}
\affiliation{CEITEC BUT, Brno University of Technology, 61200 Brno, Czech Republic}
\affiliation{Institute of Physical Engineering, Brno University of Technology, 61669 Brno, Czech Republic}
\author{M.~R\"ossle}
\affiliation{Helmholtz-Zentrum Berlin f\"ur Materialien und Energie GmbH, Wilhelm-Conrad-R\"ontgen Campus, BESSY II, 12489 Berlin, Germany}
\author{M.~Bargheer}
\email{bargheer@uni-potsdam.de}
\affiliation{Institut f\"ur Physik und Astronomie, Universit\"at Potsdam, 14476 Potsdam, Germany}
\affiliation{Helmholtz-Zentrum Berlin f\"ur Materialien und Energie GmbH, Wilhelm-Conrad-R\"ontgen Campus, BESSY II, 12489 Berlin, Germany}

\date{\today}

\begin{abstract}
By ultrafast x-ray diffraction we show that the laser-induced magnetostructural phase transition in FeRh nanoislands proceeds faster and more complete than in continuous films. We observe an intrinsic $8\,\text{ps}$ timescale for the nucleation of ferromagnetic (FM) domains \textcolor{black}{in the optically excited fraction of} both types of samples. For the continuous film, the substrate-near regions are not directly exposed to light and are only slowly transformed to the FM state \textcolor{black}{after heating above the transition temperature via near-equilibrium heat transport.} Numerical modeling of the absorption in the investigated nanoislands reveals a strong plasmonic contribution near the FeRh/MgO interface. \textcolor{black}{The larger absorption and the more homogeneous optical excitation of the nanoislands along the out-of-plane direction enables a rapid} phase transition throughout the entire volume at the intrinsic nucleation timescale.
\end{abstract}

\maketitle
\section{Introduction}
Reducing the structure-size of metallic ferromagnets to the nanoscale not only helps increasing the information storage density but also enables direct plasmonic coupling of light to the magnetic nano-bit for magnetoplasmonic control and readout\bracite{temnov2010,arme2013}. This is a particularly exciting perspective in the context of femtosecond optomagnetism\bracite{boss2016} with ultrafast optical manipulation of magnetic properties such as a polarization control of two magnetic nanolayers mediated by plasmon-polaritons \bracite{igna2019} and plasmonic enhanced all-optical switching in magnetic nanodisks\bracite{verg2023,cheng2020}. Heat assisted magnetic recording (HAMR)\bracite{rott2006, well2016} already uses \textcolor{black}{plasmonic near fields to confine the optical energy to sub-wavelength areas for highly local magnetic switching in the new generation of magnetic hard drives}. In nano-granular FePt films constituting the classical HAMR material recent experiments confirm a plasmonically enhanced ultrafast switching \bracite{gran2017}.

The potential consequences of nanostructuring FeRh go well beyond plasmonic coupling. Lateral nanostructuring limits the number of independent nucleation sites through the antiferromagnetic-to-ferromagnetic (AF-FM) phase transition around $370\,\text{K}$, which changes the nature of magnetization reversal from multi-domain to single-domain and results in discrete avalanche-like jumps of the order parameter upon cooling \bracite{uhli2016, arre2018}. In thermal equilibrium, the phase transition that is accompanied by a $1\%$ volume expansion, crucially depends on the lattice structure. The tetragonal distortion of the unit cell originating from an in-plane substrate-induced compression enhances the transition temperature \bracite{arre2018, arre2020, cheri2014}. In FeRh nanoislands, the partial relaxation of this tetragonal distortion reduces the transition temperature \bracite{arre2018, moty2023}. Generally, in-plane nano-structuring unlocks in-plane expansion on the picosecond timescale in contrast to the exclusive out-of-plane expansion of laser-excited continuous thin films \bracite{matt2023a}. The three-dimensional nature of the picosecond strain response of nanoislands preserves bulk-like material-specific expansion properties and results in a complex strain response due to cross-talk of in- and out-of-plane expansion via the Poisson effect \bracite{repp2018, repp2020, reid2018, chan2015}.

Previous experiments studied the laser-induced phase transition in FeRh by the emergence of magnetization\bracite{ju2004, berg2006, radu2010, li2022}, changes in the electronic structure\bracite{pres2021}, spin currents injected into adjacent metals\bracite{kang2023} and the rise of the larger FM lattice constant\bracite{mari2012, qui2012, matt2023b}. Probing the structural order parameter by ultrafast x-ray diffraction (UXRD), we recently disentangled optically induced nucleation and heat-transport driven propagation of the FM phase in inhomogeneously excited FeRh continuous films, whose thickness exceeds the optical penetration depth\bracite{matt2023b}. We identified a universal $8\,\text{ps}$ nucleation timescale in FeRh, which does not depend on the film thickness and temperature nor on applied laser fluence and magnetic field \bracite{matt2023b}. The effects of nanostructuring on the coupled ultrafast dynamics of demagnetization\bracite{reid2018}, remagnetization \bracite{will2019} and strain\bracite{repp2018, repp2020, reid2018} have been thoroughly studied for FePt. Ultrafast experiments on FeRh nanoislands that study the influence of the in-plane expansion, reduced number of nucleation sites and plasmonic excitation are lacking up to now.

\textcolor{black}{Here, we use UXRD to }explore the kinetics of the laser-driven phase transition in FeRh nanoislands by probing the \textcolor{black}{ultrafast emergence} of a larger lattice constant that parameterizes the FM phase as structural order parameter. In order to access the effect of finite lateral dimensions, we compare the results to a similarly thick continuous FeRh film as reference. In the nanoislands, the AF-FM phase transition drives a partial in-plane expansion of the nanoislands both in equilibrium and on ultrafast timescales. Upon laser excitation, we observe the same $8\,\text{ps}$ nucleation timescale in both samples indicating an intrinsic property of the optically induced phase transition irrespective of the sample morphology. However, while we observe a relatively slow heat transport-driven growth of the FM phase into the depth of the continuous film, the phase transition of the nanostructured film occurs precisely on the intrinsic timescale of domain nucleation. By modeling the absorption of the nanostructures, we relate this acceleration of the phase transition to a homogeneous optical excitation along the out-of-plane dimension due to plasmonic effects enabled by the size of the metallic islands below the excitation wavelength.

\section{AF-to-FM phase transition in equilibrium}
\begin{figure}[t!]
\centering
\includegraphics[width = \columnwidth]{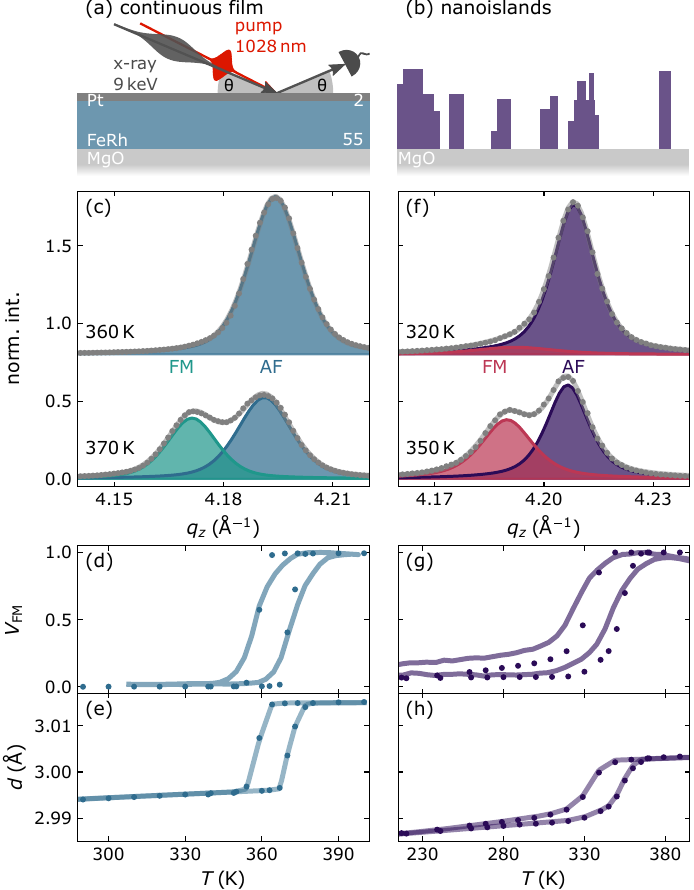}
\caption{\label{fig:fig_1_characterization} \textbf{Morphology-dependent phase transition in thermal equilibrium:} Schematic of the UXRD experiment mapping the reciprocal space via $\theta-2\theta$-scans and the sample structure of the continuous film (a) and a \textcolor{black}{sketch} of the nanoislands (b). Panels (c--e) and (f--h) characterize the equilibrium AF-FM phase transition in the continuous film and the nanostructures, respectively. (c, f) The diffracted intensity (grey symbols) is the superposition of an AF and an arising FM Bragg peak at a larger out-of-plane lattice constant during heating above $T_\text{T}$. (d, g) Temperature-dependent ferromagnetic volume fraction $V_\text{FM}$ determined by the relative integrated intensity of the FM Bragg peak (symbols) as structural order parameter and the magnetization normalized to its maximum value as magnetic order parameter (solid lines). (e, h) Temperature-dependent lattice constant (symbols) modeled by Eq.~\eqref{eq:eq_1_strain_model} using bulk expansion parameters (solid lines).}
\end{figure}

\textcolor{black}{In the first step, we characterize the morphology dependence of the AF-FM phase transition of FeRh in thermal equilibrium as a reference for the time-resolved experiments. Figure~\ref{fig:fig_1_characterization} compares the properties of the phase transition of a continuous $55\,\text{nm}$ thick FeRh(001) film with a lateral nanostructured film with a mean height of $52\,\text{nm}$.} Figures~\ref{fig:fig_1_characterization}(a) and (b) \textcolor{black}{schematically} sketch the sample structures grown on MgO(001) substrates.

Figures~\ref{fig:fig_1_characterization}(c) and (f) displays the diffracted intensity around the (002) FeRh Bragg peak along the out-of-plane reciprocal space coordinate $q_z$ recorded at the KMC-3 XPP endstation at BESSY II in the low-alpha operation mode\bracite{roes2021} with monochromized $9\,\text{keV}$ hard x-ray photons. With increasing temperature, we observe the emergence of an additional Bragg peak at smaller $q_z$ values that correspond to the arising FM phase when the temperature approaches the mean transition temperature of the thin film ($370\,\text{K}$) and the nanoislands ($350\,\text{K}$), respectively. The integrated intensity of this Bragg peak is directly proportional to the volume of FeRh exhibiting the FM phase\bracite{mari2012} and thus parameterizes the FM phase during the temperature-driven AF-FM phase transition. The proportion of the FM Bragg peak in the total intensity yields the temperature-dependent FM volume fraction $V_\text{FM}$. Figures~\ref{fig:fig_1_characterization}(d) and (g) compare this structural parameterization of the phase transition (symbols) to the macroscopic magnetization normalized to its maximum (solid lines) serving as complementary order parameter of the FM phase. Note, the heterogeneous transition temperature at different sample sites broadens the thermal hysteresis of the magnetization with respect to the locally probed structural order parameter.

The comparison of the two samples reveals a dependence of the AF-FM phase transition in thermal equilibrium on the sample morphology. The enhanced surface-to-volume ratio of the nanoislands results in a noticeable residual FM phase that persists below the transition temperature $T_\text{T}$ at the symmetry breaking surface \bracite{pres2016} and at the film-substrate interface \bracite{fan2010}. In addition, the small lateral extent of the islands partially relaxes the substrate-induced tetragonal distortion of FeRh, which lowers the transition temperature for the nanoislands \bracite{arre2018, arre2020, moty2023}. This is indicated by the lower mean out-of-plane lattice constant $d$ with respect to the continuous film (see Figs.~\ref{fig:fig_1_characterization}(e) and (h)) given by the center-of-mass (COM) of the diffracted intensity via $d=4\pi/q_{z,\text{COM}}$. This applies in particular to the out-of-plane expansion associated with the phase transition. While we find $0.4\%$ expansion for the nanoislands close to the bulk value of $0.3\,\%$\bracite{zsol1967}, the substrate-induced clamping of the in-plane expansion suppresses the Poisson effect\bracite{matt2023a} and results in an out-of-plane expansion of $0.6\%$ for the thin film. We quantified this morphology-dependent in-plane clamping by modelling the temperature-dependent lattice constant in Figs.~\ref{fig:fig_1_characterization}(e) and (h) (symbols). Utilizing literature values for the thermal expansion of FeRh and MgO, Equation~\eqref{eq:eq_1_strain_model} (solid lines) yields excellent agreement if the thin film purely follows the in-plane expansion of the MgO substrate and if $58\,\%$ of the volume of the nanoislands behave bulk-like where the relaxation of the in-plane constraints is expected to increase towards the surface and to depend on the in-plane dimensions of the different nanoislands \bracite{lask2019} (see Method section).

\section{Ultrafast phase transition tracked by UXRD}
\begin{figure}[t!]
\centering
\includegraphics[width = \columnwidth]{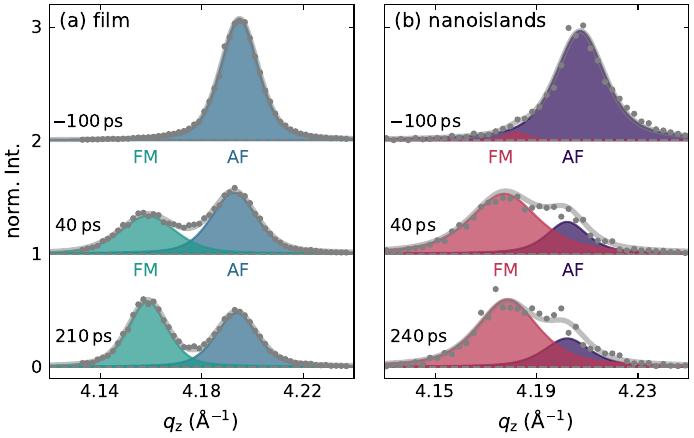}
\caption{\label{fig:fig_2_rocking} \textbf{\textcolor{black}{Morphology-dependent tetragonal distortion across the laser-induced phase transition}:} (a) Transient Bragg peaks of the thin film for an excitation of $11.2\,\text{mJ/cm}^2$ at $340\,\text{K}$ dissected into the FM (green) and AF (blue) Bragg peak that are well-separated in reciprocal space. (b) In the nanoislands, the FM (pink) and AF (purple) Bragg peak are less separated due to the partial in-plane expansion of the unit cell across the laser-induced phase transition for $5.2\,\text{mJ/cm}^2$ at $230\,\text{K}$. The data for different pump-probe delays are vertically off-set to improve visibility.}
\end{figure}

In the UXRD experiment, we track the emergence of the FM Bragg peak upon excitation by a $600\,\text{fs}$ near-infrared laser pulse. The time-dependent integral of the FM Bragg peak quantifies the transient laser-induced FM volume fraction $V_\text{FM}$. Figures~\ref{fig:fig_2_rocking}(a) and (b) display the diffracted intensity along $q_z$ before and after an excitation with $11.2\,\text{mJ/cm}^2$ at $340\,\text{K}$ for the thin film and with $5.2\,\text{mJ/cm}^2$ at $230\,\text{K}$ for the nanoislands, respectively. The emerging FM Bragg peaks indicate the optically induced AF-FM phase transition for both samples. The AF and FM Bragg peaks are well separated for the thin film. For the nanoislands, an ultrafast in-plane expansion on short timescales is enabled by their small lateral extent\bracite{matt2023a}. The concomitant transient out-of-plane Poisson-contraction results in less separated Bragg peaks (Fig.~\ref{fig:fig_2_rocking}(b)) for the nanoislands. This indicates a reduced tetragonal distortion of the unit cell upon laser-excitation and the emergence of more cubic FM unit cells upon nucleation in the nanoislands as already discussed for thermal equilibrium. In addition to the lattice constant change across the phase transition, the dynamics of the laser-induced phase transition also depends on the sample morphology. While the integrated intensity of the FM Bragg peak barely changes between $40$ and $240\,\text{ps}$ for the nanoislands, the FM Bragg peak clearly increases after $40\,\text{ps}$ for the thin film.

\begin{figure}[t!]
\centering
\includegraphics[width = \columnwidth]{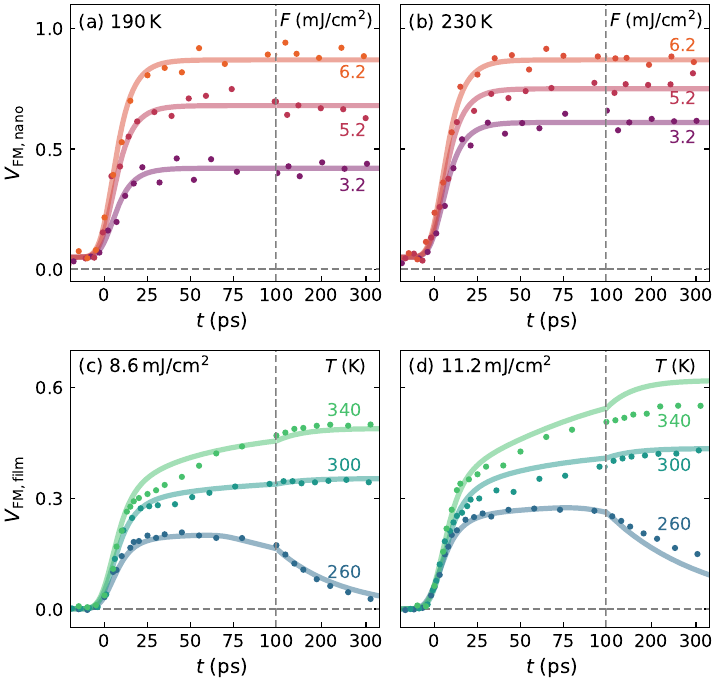}
\caption{\label{fig:fig_3_transient_FM} \textbf{\textcolor{black}{Morphology-dependent dynamics of the laser-induced FM volume fraction}:} (a) Transient FM volume fraction of the nanoislands at $T=190\,\text{K}$ for various fluences $F$. (b) Same for $T=230\,\text{K}$, which increases the conversion to the FM state at low fluence. (c) Temperature series at a relatively low fluence of $F=8.6\,\text{mJ/cm}^2$ for the thin film. (d) Same for $F=11.2.0\,\text{mJ/cm}^2$. The two-step rise of $V_\text{FM}$ in the thin film (c and d) indicates propagation of the nucleated FM phase into the depth of the layer driven by near-equilibrium heat transport. In (a) and (b), solid lines denote the kinetics of FM domain nucleation according to Eq.~\eqref{eq:eq_3_exponential} convoluted with the time-resolution given by the duration of the x-ray pulse. \textcolor{black}{The solid lines in (c) and (d) represent the modelled $V_\text{FM}(t)$ considering the time-dependent heating above $T_\text{AF-FM}$ via heat diffusion (see Sec.~\ref{sec:methods} for further details). The vertical dashed lines denote a break in the delay axis.}}
\end{figure}

Figure~\ref{fig:fig_3_transient_FM} displays the resulting transient FM volume fraction $V_\text{FM}$ for both samples under various excitation conditions (symbols). \textcolor{black}{For the nanoislands, we observe an exponential rise of $V_\text{FM}$ associated with an optically induced nucleation of FM domains \bracite{mari2012,matt2023b} according to:}
\begin{align}
    V_\text{FM}(t) &= V_\text{FM}^* \cdot \left( 1-e^{-t/\tau} \right) \;,
\label{eq:eq_3_exponential}
\end{align}
\textcolor{black}{with the universal nucleation timescale $\tau=8\,\text{ps}$ previously identified for thin films\bracite{matt2023b}. The convolution of the exponential growth according to Eq.~\eqref{eq:eq_3_exponential} with the $17\,\text{ps}$-long x-ray pulse limiting the time-resolution (solid lines) yields an excellent agreement with the experimental $V_\text{FM}(t)$ of the nanoislands in Figs.~\ref{fig:fig_3_transient_FM}(a) and (b).} The final FM volume fraction $V_\text{FM}^*$ is adjusted to the experimental value of $V_\text{FM}(t=40\,\text{ps})$ for the respective measurement and we include the residual FM phase in Fig.~\ref{fig:fig_1_characterization}(g) being present before excitation. With increasing fluence and initial sample temperature a larger fraction of the nanoislands volume is excited above the critical threshold characteristic for the first-order phase transition\bracite{berg2006, mari2012}, which results in an enhanced $V_\text{FM}^*$.

\textcolor{black}{For the continuous thin film, we observe a strong dependence of the rising $V_\text{FM}$ on the initial sample temperature (see Figs.~\ref{fig:fig_3_transient_FM}(c) and (d)). At $260\,\text{K}$ the FM volume fraction rises on the same timescale as in the nanoislands and reaches its maximum within $40\,\text{ps}$. For higher temperatures, we observe an additional slow rise of $V_\text{FM}$ after $40\,\text{ps}$. The amplitude of this additional contribution as well as its timescale depend on the fluence and temperature. We recently associated this delayed rise with a propagation of the FM phase into the depth of the inhomogeneously excited film (see Fig.~\ref{fig:fig_4_absorption}(f)) via near-equilibrium heat transport that heats the substrate-near part above the transition temperature\bracite{matt2023b}. Additionally, we showed that the propagation of the FM phase is slower than expected from the equilibrium heat transport. Here, we nicely capture the experimental fluence- and temperature-dependent $V_\text{FM}(t)$ by a straight forward model (solid lines in Figs.~\ref{fig:fig_3_transient_FM}(c) and (d)),\textcolor{black}{which keeps the nucleation timescale fixed in all graphs of Figs.~\ref{fig:fig_3_transient_FM}. The model describing additional dynamics of $V_\text{FM}$ in in Figs.~\ref{fig:fig_3_transient_FM}c) and d) beyond $40\,\text{ps}$ considers the heating of FeRh above the transition temperature calculated by solving the 1-D } heat diffusion equation using literature values of the thermo-physical parameters. Further details are given in the methods section~\ref{sec:methods}. The excellent simultaneous agreement with both the fluence- and temperature-dependent rising dynamics and amplitudes with a fixed set of parameters verifies our interpretation of a propagation of the FM phase into the depth of the inhomogeneously excited thin film via heat diffusion.}

\textcolor{black}{In addition to this morphology-dependent dynamics of $V_\text{FM}$, Figs.~\ref{fig:fig_3_transient_FM}(a) and (b) show that a fluence of $6.2\,\text{mJ/cm}^2$ induces the FM phase in nearly the complete volume of the FeRh nanoislands within $40\,\text{ps}$. In contrast, Figs.~\ref{fig:fig_3_transient_FM}(c) and (d) show that nearly twice the fluence only transforms $35\%$ of the continuous thin film within $40\,\text{ps}$. After $40\,\text{ps}$ the FM volume fraction continues to rise slowly, however not beyond $60\%$. This comparison unambiguously demonstrates that the laser-induced AF-FM phase transition in laterally nanostructured FeRh is faster and more complete compared to the continuous thin film. This observation cannot be explained by the additional $2\,\text{nm}$ Pt capping layer of the continuous film that only reduces the absorption in FeRh by $\approx 10\,\%$.}

\section{Modeling the absorbed light power by classical electrodynamics}
\textcolor{black}{To explain the observed morphology dependence of the laser-induced AF-FM phase transition, we calculate the local energy dissipation in the nanoislands by classical electrodynamics using the software package \textsc{COMSOL} Multiphysics. To describe the experiment as realistically as possible} we reproduce the topography of the nanoislands characterized by atomic force microscopy (AFM) (Fig.~\ref{fig:fig_4_absorption}(a)) in \textsc{COMSOL} by ellipsoids utilizing an algorithm for rapid contour detection \bracite{rasc2018} (see Fig.~\ref{fig:fig_4_absorption}(b)).

Figure~\ref{fig:fig_4_absorption}(c) displays the calculated local power absorption $P_\text{abs}$ of the nanostructures for the excitation conditions in the experiment\textcolor{black}{, i.e., p-polarized light with a wavelength of $1028\,\text{nm}$ incident under $60\,^\circ$ with respect to the sample normal} utilizing the refractive index of MgO \bracite{step1952} $n_\text{MgO}=1.72$ and of FeRh $n_\text{FeRh}=4.67+5.58i$ measured via spectroscopic ellipsometry at the pump laser wavelength $\lambda =1028\,\text{nm}$. The calculated local power absorption $P_\text{abs}$ reveals the existence of hot-spots with drastically enhanced absorption. By fitting an exponential decay function to the local $z$-dependent absorption (FeRh-MgO interface corresponds to $z=0$), we find a large spread of the optical penetration depth $\delta_\text{p}$. Figure~\ref{fig:fig_4_absorption}(d) shows this distribution relative to the semi-infinite medium value $\delta_{\text{p},0}=14.7\,\text{nm}$, \textcolor{black}{that also applies to the continuous film}. Yellow color-code indicates a locally strongly enhanced optical penetration depth due to nanostructuring that leads to a more homogeneous excitation along $z$. Figure~\ref{fig:fig_4_absorption}(e) depicts an exemplary $z$-dependent cross section of the absorption as function of the in-plane $x$ coordinate at $y=3.2\,\text{\textmu m}$. This lineout displays an enhanced absorption near the FeRh-MgO interface in several nanoislands indicated by the yellow color.

To compare these results to a continuous film, we determine the average total absorption in the nanoislands as function of the distance from the FeRh-MgO interface ($z=0$) from the local power absorption $P_\text{abs}(x,y)$ in Fig.~\ref{fig:fig_4_absorption}(c). Figure~\ref{fig:fig_4_absorption}(f) displays the average $z$-dependent absorption of the nanoislands (symbols) and of a $55\,\text{nm}$ continuous FeRh film scaled by the surface coverage of the nanoislands ($49\,\%$) in the studied sample (grey solid line). This comparison highlights a more homogeneous excitation of the nanoislands and a strong enhancement of the absorption in the substrate-near region that is barely excited in the continuous thin film despite the comparable average height of the nanoislands. Comparing the $z$-integrated absorption of the nanoislands and the continuous film in Fig.~\ref{fig:fig_4_absorption}(f), we find that the total optical absorption of the nanostructures amounts to $34\,\%$ of the incident power, which exceeds the absorption of the continuous FeRh film by a factor of $1.5$.

\section{Plasmonic absorption}
\begin{figure}[t!]
\centering
\includegraphics[width = \columnwidth]{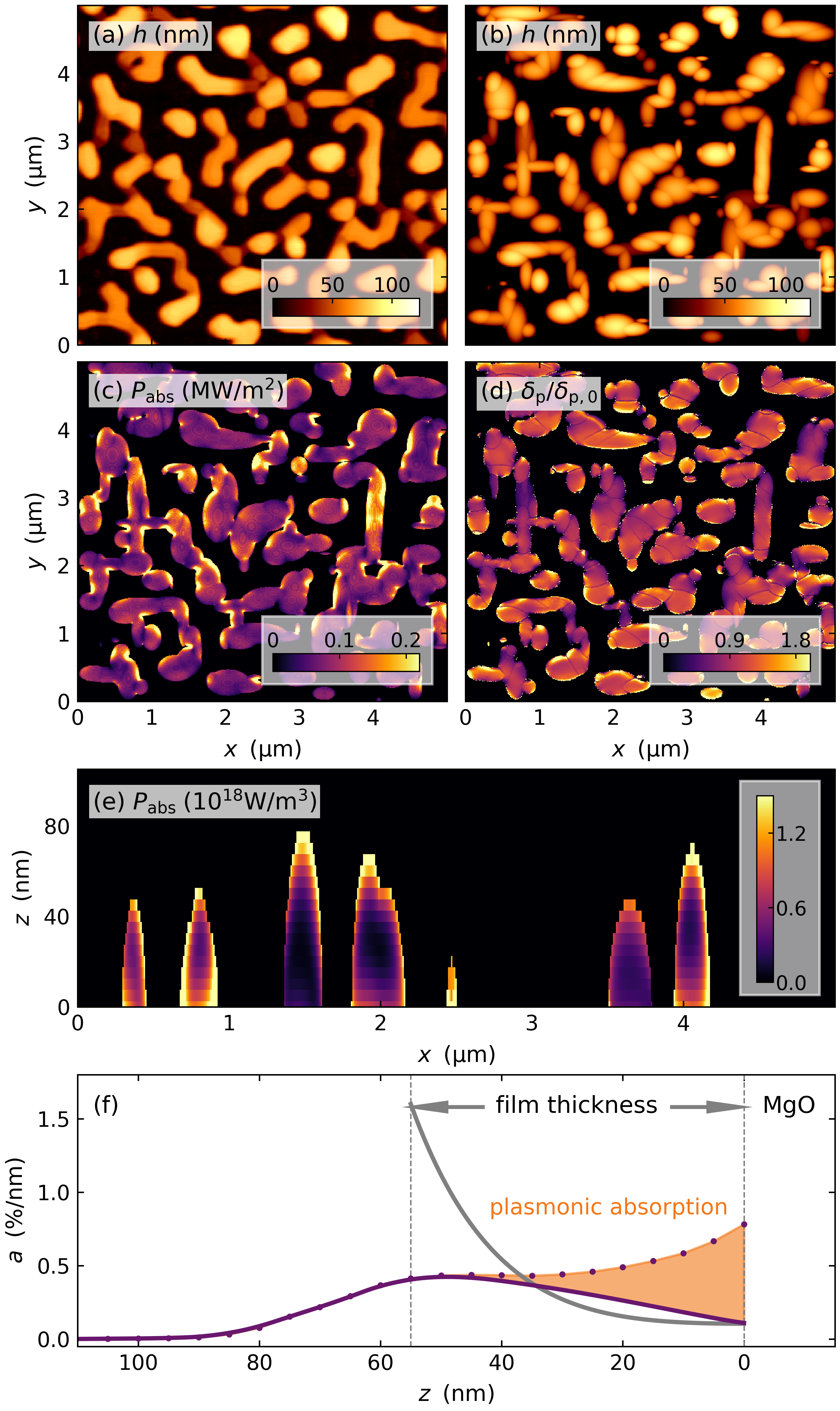}
\caption{\label{fig:fig_4_absorption} \textbf{\textcolor{black}{Depth-dependent} optical excitation of nanostructured FeRh:} Re-build of the topography of the FeRh nanostructures characterized by AFM (a) in \textsc{COMSOL} (b). \textcolor{black}{The AFM data are the same as in our previous publication\bracite{moty2023} since we use the identical sample in this study}. (c) Local absorbed power per area of the FeRh nanostructures calculated in \textsc{COMSOL} by solving the Maxwell equations for $\lambda=1028\,\text{nm}$. (d) Local optical penetration depth determined from $P_\text{abs}$ as function of the depth relative to the local height $h$. (e) Absorption of different nanoislands as function of $z$ at $y=3.2\,\text{\textmu m}$. (f) Total absorption profile integrated over all pixels (purple symbols). The purple solid line corresponds to the hypothetical scenario of $z$-dependent absorption integrated over all pixels, for each assuming an absorption profile of a continuous film of an equivalent thickness given by $\delta_{\text{p},0}$. The grey line denotes the absorption profile of the continuous $55\,\text{nm}$ thick FeRh film. \textcolor{black}{The orange shaded area represents the effect of the plasmonic absorption in the nanoislands.}}
\end{figure}

\textcolor{black}{The enhanced absorption near the FeRh-MgO interface\bracite{garc2022} responsible for the local absorption hot-spots shown in Fig.~\ref{fig:fig_4_absorption}(c) is characteristic of localized surface plasmon polaritons (LSPP).\bracite{kreibig2013,maier2007} The associated resonance in the absorption spectrum crucially depends on the spatial dimensions of the nanostructures. Changing the diameter of metal nanodiscs with constant height shifts the resonance across the entire visible and near-infrared spectrum for various metals.\bracite{zoric2011} An additional red shift and concomitant broadening is observed when changing the aspect ratio of ellipsoids at constant volume \bracite{kelly2003}. The distribution of nanoislands in our FeRh sample is both inhomogeneous in size and aspect ratio, which leads to an ultrabroad featureless absorption spectrum. Nonetheless, the plasmonic absorption mediated by the nanoscale size of a metal and characterized by a negative permittivity (see Fig.~\ref{fig:fig_5_refractive_index}) is central to explain the large amount of light energy dissipated near the FeRh-MgO interface, i.e.} at depths that receive only a negligible amount of light energy in continuous thin films.

\textcolor{black}{The calculated $z$-dependent absorption of the nanoislands in Fig.~\ref{fig:fig_4_absorption}(f) contains signatures of both the dissipation of light energy via plasmons and the local FeRh height that relates a single height $z$ to different depth relative to the local surface. In order to isolate the contribution from the plasmonic absorption, we calculate the $z$-dependent absorption of the nanoislands sample by assuming that the absorption of each pixel is identical to that of a continuous film of a thickness equivalent to the local FeRh height. Thus, the $z$-dependent absorption for each pixel is determined by the semi-infinite medium value of the optical penetration depth $\delta_{\text{p},0}$ starting at the local height. The sum over all pixels yields the purple solid line in Fig.~\ref{fig:fig_4_absorption}(f) that matches the \textsc{COMSOL} simulation for $z>50\,\text{nm}$ where the absorption decreases with increasing $z$ because \textcolor{black}{only nanoislands higher than the $z$ value contribute to absorption.} However, the average absorption of nanostructures (symbols) for $z<30\,\text{nm}$ is much larger than that predicted by the pixel-integrated absorption of equivalent thin films, which neglects the plasmonic enhancement near the FeRh-MgO interface.}

\textcolor{black}{This difference highlighted by the orange shaded area originates from the plasmonic absorption that increases the absorption beyond integration of a local Lambert-Beer exponential decay. It enhances the absorption near the FeRh-MgO interface where the plasmonic absorption is particularly strong because the dielectric response of the substrate enhances the field. This additional dissipation of the light energy in the nanoislands leads to a more homogeneous optical excitation along the $z$ direction with respect to continuous thin films. Additionally, Fig.~\ref{fig:fig_4_absorption}(f) displays that the plasmonic absorption significantly increases the total absorption of the nanoislands.}

This stronger and almost homogeneous excitation of the complete volume of the nanostructures supports the $8\,\text{ps}$ nucleation-driven phase transition throughout the entire nanoislands. This suppresses the slow phase transition via near-equilibrium heat transport observed for the thin film (Figs.~\ref{fig:fig_3_transient_FM}(c--d)) as the majority of the nanoislands' volume is optically excited. Therefore, the laser-induced phase transition in FeRh nanostructures with small lateral extension is drastically accelerated and even more efficiently driven due to the overall enhanced absorption.

\section{Conclusion}
In summary, we studied the morphology dependence of the laser-induced AF-FM phase transition by comparing a continuous and a nanostructured thin FeRh film. We find an ultrafast in-plane expansion of the nanoislands, whereas the thin FeRh film is pinned to the MgO. This results in a less tetragonal distortion of the unit cell across the phase transition, however, it has no influence on the nucleation timescale of the FM domains. Instead, only the \textcolor{black}{change of the absorption profile due to plasmons} affects the dynamics of the phase transition: By modelling the spatially resolved optical absorption of the FeRh nanostructures, we identified an enhanced absorption near the FeRh-MgO interface and an enhanced optical penetration depth. This results in a homogeneous excitation of the nanoislands, which drives a nucleation of FM domains on an $8\,\text{ps}$ timescale within the volume of the FeRh nanostructures and makes slow heat-transport driven propagation of the FM phase irrelevant. This accelerates the phase transition in comparison with the continuous film that exhibits nucleation only within the optically excited near-surface region and shows a subsequent slow growth of the FM phase into the depth of film at initial sample temperatures slightly below the transition temperature.

\section{Methods} \label{sec:methods}
\begin{small}
\textit{Sample growth and characterization:} The continuous $55\,\text{nm}$ thick FeRh(001) film was grown by magnetron sputtering from an equiatomic FeRh target \bracite{arre2020} and capped by $2\,\text{nm}$ of Pt. The nanostructured sample is composed of epitaxial FeRh(001) nanoislands formed by solid state dewetting of an $20\,\text{nm}$-thick epitaxial FeRh(001) film via self-assembly resulting in maze-like structures\bracite{moty2023} with a mean height of $52\,\text{nm}$. \textcolor{black}{The sample is the same as in our previous publication that also shows larger area scanning electron microscopy images\bracite{moty2023}.} The magnetization is measured via vibrating sample magnetometry using a QuantumDesign VersaLab magnetometer. \textcolor{black}{AFM topography measurements were performed using a Dimension Icon microscope from Bruker Corporation while employing commercial MESP probes. Figure~\ref{fig:fig_5_refractive_index} displays the complex dielectric function of a $30\,\text{nm}$-thick continuous FeRh film as function of the wavelength $\lambda$ experimentally determined via ellipsometry. Literature consistently reports a negative value of the real part $\epsilon_1$ for the visible an near infrared, with spectral details depending on sample preparation.\cite{chen1988, rhee1995, saidl2016}}
\begin{figure}[h!]
\centering
\includegraphics[width = 0.8\columnwidth]{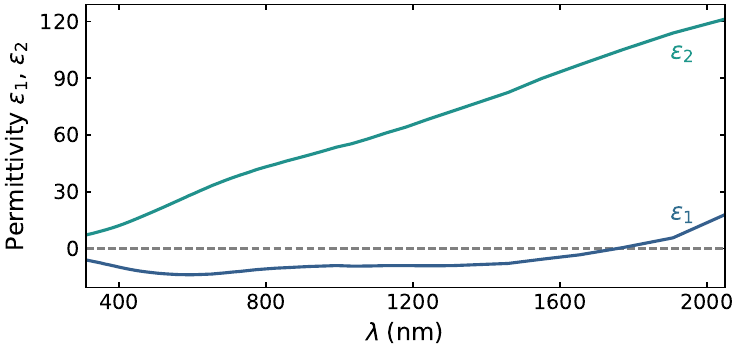}
\caption{\label{fig:fig_5_refractive_index} \textbf{Spectral ellipsometry results for the complex dielectric function $\epsilon_1+i\epsilon_2$} of a $30\,\text{nm}$-thick continuous FeRh film. The negative $\epsilon_1$ for wavelength between $300$ and $1800\,\text{nm}$ demonstrates that FeRh supports LSPRs for a borad range of excitation wavelength around the pump wavelength of $1028\,\text{nm}$ used in the UXRD experiment.}
\end{figure}

\textit{UXRD experiment:} The FeRh samples are excited by $600\,\text{fs}$ p-polarized pump pulses with a central wavelength of $1028\,\text{nm}$ and a repition reate of $100\,\text{kHz}$ that are incident at $\approx 30^\circ$ with respect to the sample surface as sketched in Fig.~\ref{fig:fig_1_characterization}(a). \textcolor{black}{The footprint of the pump pulse on the sample was $1300 \times 900\,\text{\textmu m}^2$ (major and minor FWHM-diameter), which is $9$ times larger than the footprint of the x-ray probe pulse ($400 \times 300\,\text{\textmu m}^2$) to ensure a laterally homogeneous excitation of the probed volume.} We probe the emergence of the FM Bragg peak that parameterizes the laser-induced AF-FM phase transition by $9\,\text{keV}$ hard x-ray pulses with a pulse duration of $17\,\text{ps}$\bracite{roes2021} performing symmetric $\theta$-$2\theta$ scans around the (002) Bragg reflection at $28^\circ$. The diffracted x-rays are recorded on an area detector (Dectris PILATUS 100K).

\textit{Modelling $d(T)$ in equilibrium:} The influence of the substrate-induced in-plane clamping on the out-of-plane expansion of the FeRh samples is described by \bracite{matt2023a}:
\begin{align}
	\alpha_\text{eff} &= \alpha_\text{bulk}(T) + 2 \chi \frac{c_{1133}}{c_{3333}} \left( \alpha_\text{bulk}(T) - \alpha_\text{MgO} \right) \; ,
\label{eq:eq_1_strain_model}
\end{align}
where $\alpha_\text{bulk}(T)$ and $\alpha_\text{MgO}=10.5\cdot10^{-6} \text{K}^{-1}$ denote the thermal expansion coefficients of bulk FeRh and MgO, respectively. The expansion of FeRh $\alpha_\text{bulk}(T)$ is given by the expansion coefficients $\alpha^\text{FM}=6.0\cdot10^{-6} \text{K}^{-1}$ and $\alpha^\text{AF}=9.7\cdot10^{-6} \text{K}^{-1}$ in the AF and FM phase\bracite{ibar1994} and the expansion of $0.3\,\%$ across the AF-FM phase transition\bracite{zsol1967}, considering the temperature-dependent volume fraction in the FM phase $V_\text{FM}(T)$, which we derived from the integrated intensity of the FM Bragg peak in Figs.~\ref{fig:fig_1_characterization}(c) and (f).  The elastic constants of FeRh $c_{1133}$ and $c_{3333}$ quantify the effect of the in-plane expansion on the out-of-plane expansion via the Poisson effect \bracite{matt2023a} and $\alpha_\text{eff}$ denotes the modified expansion coefficient of the samples depending on the parameter $\chi$. This parameter measures the epitaxy to the substrate, where $\chi=0$ corresponds to pure bulk-like in-plane expansion and $\chi=1$ to an in-plane expansion completely determined by the MgO substrate. We find excellent agreement for $\chi=1$ and $\chi=0.42$ for the continuous film and the nanoislands, respectively.

\textit{Modelling $V_\text{FM}(t)$ in the continuous film:} In order to describe the fluence- and temperature-dependent additional slow rise of $V_\text{FM}$ in Figs.~\ref{fig:fig_3_transient_FM}(c) and (d), we explicitly consider the heat transport within the inhomogeneously excited FeRh film ($\delta_{\text{p},0}=14.7\,\text{nm}$) by solving the one-dimensional heat diffusion equation utilizing literature values of the thermo-physical properties of Pt, FeRh and MgO as described previously \bracite{matt2023b}. We assume $V_\text{FM}$ to rise on the $8\,\text{ps}$ nucleation timescale in the fraction of the FeRh film that is heated above the average AF-FM transition temperature $T_\text{AF-FM}=370\,\text{K}$ (see Fig.~\ref{fig:fig_1_characterization}(e)) directly after the excitation ($t<200\,\text{fs}$). This already yields excellent agreement for the measurements at $260\,\text{K}$ within the first $50\,\text{ps}$. In order to obtain a good fit for the delayed rise of $V_\text{FM}$ by heat transport, we assume an exponential rise on a $60\,\text{ps}$ timescale after the local temperature exceeds $T_\text{AF-FM}$ at larger delays due to the heat transport. To match the decrease of $V_\text{FM}$, i.e., the recovery of the AF phase on hundreds of picoseconds, we empirically assume an exponential decay of the FM phase on a $220\,\text{ps}$ timescale, after the local temperature falls below the average FM-AF transition temperature $T_\text{FM-AF}=350\,\text{K}$ (see Fig.~\ref{fig:fig_1_characterization}(e)).
\end{small}

\section*{Acknowledgements}
We acknowledge the DFG for financial support via Project-No.\ 328545488 – TRR 227 project A10 and the BMBF for funding via 05K22IP1. Access to the CEITEC Nano Research Infrastructure was supported by the Ministry of Education, Youth and Sports (MEYS) of the Czech Republic under the project CzechNanoLab (LM2023051). Measurements were carried out at the KMC3-XPP instrument at the BESSY II electron storage ring operated by the Helmholtz-Zentrum Berlin für Materialien und Energie.
\bibliographystyle{achemso}
\bibliography{references.bib}

\providecommand{\latin}[1]{#1}
\makeatletter
\providecommand{\doi}
  {\begingroup\let\do\@makeother\dospecials
  \catcode`\{=1 \catcode`\}=2 \doi@aux}
\providecommand{\doi@aux}[1]{\endgroup\texttt{#1}}
\makeatother
\providecommand*\mcitethebibliography{\thebibliography}
\csname @ifundefined\endcsname{endmcitethebibliography}
  {\let\endmcitethebibliography\endthebibliography}{}
\begin{mcitethebibliography}{45}
\providecommand*\natexlab[1]{#1}
\providecommand*\mciteSetBstSublistMode[1]{}
\providecommand*\mciteSetBstMaxWidthForm[2]{}
\providecommand*\mciteBstWouldAddEndPuncttrue
  {\def\EndOfBibitem{\unskip.}}
\providecommand*\mciteBstWouldAddEndPunctfalse
  {\let\EndOfBibitem\relax}
\providecommand*\mciteSetBstMidEndSepPunct[3]{}
\providecommand*\mciteSetBstSublistLabelBeginEnd[3]{}
\providecommand*\EndOfBibitem{}
\mciteSetBstSublistMode{f}
\mciteSetBstMaxWidthForm{subitem}{(\alph{mcitesubitemcount})}
\mciteSetBstSublistLabelBeginEnd
  {\mcitemaxwidthsubitemform\space}
  {\relax}
  {\relax}

\bibitem[Temnov \latin{et~al.}(2010)Temnov, Armelles, Woggon, Guzatov,
  Cebollada, Garcia-Martin, Garcia-Martin, Thomay, Leitenstorfer, and
  Bratschitsch]{temnov2010}
Temnov,~V.~V.; Armelles,~G.; Woggon,~U.; Guzatov,~D.; Cebollada,~A.;
  Garcia-Martin,~A.; Garcia-Martin,~J.-M.; Thomay,~T.; Leitenstorfer,~A.;
  Bratschitsch,~R. Active magneto-plasmonics in hybrid metal--ferromagnet
  structures. \emph{Nature Photonics} \textbf{2010}, \emph{4}, 107--111\relax
\mciteBstWouldAddEndPuncttrue
\mciteSetBstMidEndSepPunct{\mcitedefaultmidpunct}
{\mcitedefaultendpunct}{\mcitedefaultseppunct}\relax
\EndOfBibitem
\bibitem[Armelles \latin{et~al.}(2013)Armelles, Cebollada, García-Martín, and
  González]{arme2013}
Armelles,~G.; Cebollada,~A.; García-Martín,~A.; González,~M.~U.
  Magnetoplasmonics: Combining Magnetic and Plasmonic Functionalities.
  \emph{Advanced Optical Materials} \textbf{2013}, \emph{1}, 10--35\relax
\mciteBstWouldAddEndPuncttrue
\mciteSetBstMidEndSepPunct{\mcitedefaultmidpunct}
{\mcitedefaultendpunct}{\mcitedefaultseppunct}\relax
\EndOfBibitem
\bibitem[Bossini \latin{et~al.}(2016)Bossini, Belotelov, Zvezdin, Kalish, and
  Kimel]{boss2016}
Bossini,~D.; Belotelov,~V.~I.; Zvezdin,~A.~K.; Kalish,~A.~N.; Kimel,~A.~V.
  Magnetoplasmonics and Femtosecond Optomagnetism at the Nanoscale. \emph{ACS
  Photonics} \textbf{2016}, \emph{3}, 1385--1400\relax
\mciteBstWouldAddEndPuncttrue
\mciteSetBstMidEndSepPunct{\mcitedefaultmidpunct}
{\mcitedefaultendpunct}{\mcitedefaultseppunct}\relax
\EndOfBibitem
\bibitem[Ignatyeva \latin{et~al.}(2019)Ignatyeva, Davies, Sylgacheva,
  Tsukamoto, Yoshikawa, Kapralov, Kirilyuk, Belotelov, and Kimel]{igna2019}
Ignatyeva,~D.; Davies,~C.; Sylgacheva,~D.; Tsukamoto,~A.; Yoshikawa,~H.;
  Kapralov,~P.; Kirilyuk,~A.; Belotelov,~V.; Kimel,~A. Plasmonic
  layer-selective all-optical switching of magnetization with nanometer
  resolution. \emph{Nature communications} \textbf{2019}, \emph{10}, 4786\relax
\mciteBstWouldAddEndPuncttrue
\mciteSetBstMidEndSepPunct{\mcitedefaultmidpunct}
{\mcitedefaultendpunct}{\mcitedefaultseppunct}\relax
\EndOfBibitem
\bibitem[Verg{\`e}s \latin{et~al.}(2023)Verg{\`e}s, Perumbilavil, Hohlfeld,
  Freire-Fern{\'a}ndez, Le~Guen, Kuznetsov, Montaigne, Malinowski, Lacour,
  Hehn, \latin{et~al.} others]{verg2023}
Verg{\`e}s,~M.; Perumbilavil,~S.; Hohlfeld,~J.; Freire-Fern{\'a}ndez,~F.;
  Le~Guen,~Y.; Kuznetsov,~N.; Montaigne,~F.; Malinowski,~G.; Lacour,~D.;
  Hehn,~M.; others Energy Efficient Single Pulse Switching of [Co/Gd/Pt] N
  Nanodisks Using Surface Lattice Resonances. \emph{Advanced Science}
  \textbf{2023}, \emph{10}, 2204683\relax
\mciteBstWouldAddEndPuncttrue
\mciteSetBstMidEndSepPunct{\mcitedefaultmidpunct}
{\mcitedefaultendpunct}{\mcitedefaultseppunct}\relax
\EndOfBibitem
\bibitem[Cheng \latin{et~al.}(2020)Cheng, Wang, Su, Wang, Cai, Sun, and
  Liu]{cheng2020}
Cheng,~F.; Wang,~C.; Su,~Z.; Wang,~X.; Cai,~Z.; Sun,~N.~X.; Liu,~Y. All-optical
  manipulation of magnetization in ferromagnetic thin films enhanced by
  plasmonic resonances. \emph{Nano Letters} \textbf{2020}, \emph{20},
  6437--6443\relax
\mciteBstWouldAddEndPuncttrue
\mciteSetBstMidEndSepPunct{\mcitedefaultmidpunct}
{\mcitedefaultendpunct}{\mcitedefaultseppunct}\relax
\EndOfBibitem
\bibitem[Rottmayer \latin{et~al.}(2006)Rottmayer, Batra, Buechel, Challener,
  Hohlfeld, Kubota, Li, Lu, Mihalcea, Mountfield, Pelhos, Peng, Rausch,
  Seigler, Weller, and Yang]{rott2006}
Rottmayer,~R. \latin{et~al.}  Heat-Assisted Magnetic Recording. \emph{IEEE
  Transactions on Magnetics} \textbf{2006}, \emph{42}, 2417--2421\relax
\mciteBstWouldAddEndPuncttrue
\mciteSetBstMidEndSepPunct{\mcitedefaultmidpunct}
{\mcitedefaultendpunct}{\mcitedefaultseppunct}\relax
\EndOfBibitem
\bibitem[Weller \latin{et~al.}(2016)Weller, Parker, Mosendz, Lyberatos, Mitin,
  Safonova, and Albrecht]{well2016}
Weller,~D.; Parker,~G.; Mosendz,~O.; Lyberatos,~A.; Mitin,~D.; Safonova,~N.~Y.;
  Albrecht,~M. FePt heat assisted magnetic recording media. \emph{Journal of
  Vacuum Science \& Technology B} \textbf{2016}, \emph{34}\relax
\mciteBstWouldAddEndPuncttrue
\mciteSetBstMidEndSepPunct{\mcitedefaultmidpunct}
{\mcitedefaultendpunct}{\mcitedefaultseppunct}\relax
\EndOfBibitem
\bibitem[Granitzka \latin{et~al.}(2017)Granitzka, Jal, Le~Guyader, Savoini,
  Higley, Liu, Chen, Chase, Ohldag, Dakovski, Schlotter, Carron, Hoffman, Gray,
  Shafer, Arenholz, Hellwig, Mehta, Takahashi, Wang, Fullerton, Stöhr, Reid,
  and Dürr]{gran2017}
Granitzka,~P.~W. \latin{et~al.}  Magnetic Switching in Granular FePt Layers
  Promoted by Near-Field Laser Enhancement. \emph{Nano Letters} \textbf{2017},
  \emph{17}, 2426--2432\relax
\mciteBstWouldAddEndPuncttrue
\mciteSetBstMidEndSepPunct{\mcitedefaultmidpunct}
{\mcitedefaultendpunct}{\mcitedefaultseppunct}\relax
\EndOfBibitem
\bibitem[Uhl{\'\i}{\v{r}} \latin{et~al.}(2016)Uhl{\'\i}{\v{r}}, Arregi, and
  Fullerton]{uhli2016}
Uhl{\'\i}{\v{r}},~V.; Arregi,~J.~A.; Fullerton,~E.~E. Colossal magnetic phase
  transition asymmetry in mesoscale FeRh stripes. \emph{Nature communications}
  \textbf{2016}, \emph{7}, 13113\relax
\mciteBstWouldAddEndPuncttrue
\mciteSetBstMidEndSepPunct{\mcitedefaultmidpunct}
{\mcitedefaultendpunct}{\mcitedefaultseppunct}\relax
\EndOfBibitem
\bibitem[Arregi \latin{et~al.}(2018)Arregi, Hork{\'y}, Fabianov{\'a}, Tolley,
  Fullerton, and Uhl{\'\i}{\v{r}}]{arre2018}
Arregi,~J.~A.; Hork{\'y},~M.; Fabianov{\'a},~K.; Tolley,~R.; Fullerton,~E.~E.;
  Uhl{\'\i}{\v{r}},~V. Magnetization reversal and confinement effects across
  the metamagnetic phase transition in mesoscale FeRh structures. \emph{Journal
  of Physics D: Applied Physics} \textbf{2018}, \emph{51}, 105001\relax
\mciteBstWouldAddEndPuncttrue
\mciteSetBstMidEndSepPunct{\mcitedefaultmidpunct}
{\mcitedefaultendpunct}{\mcitedefaultseppunct}\relax
\EndOfBibitem
\bibitem[Arregi \latin{et~al.}(2020)Arregi, Caha, and
  Uhl{\'\i}{\v{r}}]{arre2020}
Arregi,~J.~A.; Caha,~O.; Uhl{\'\i}{\v{r}},~V. Evolution of strain across the
  magnetostructural phase transition in epitaxial FeRh films on different
  substrates. \emph{Physical Review B} \textbf{2020}, \emph{101}, 174413\relax
\mciteBstWouldAddEndPuncttrue
\mciteSetBstMidEndSepPunct{\mcitedefaultmidpunct}
{\mcitedefaultendpunct}{\mcitedefaultseppunct}\relax
\EndOfBibitem
\bibitem[Cherifi \latin{et~al.}(2014)Cherifi, Ivanovskaya, Phillips, Zobelli,
  Infante, Jacquet, Garcia, Fusil, Briddon, Guiblin, \latin{et~al.}
  others]{cheri2014}
Cherifi,~R.; Ivanovskaya,~V.; Phillips,~L.; Zobelli,~A.; Infante,~I.;
  Jacquet,~E.; Garcia,~V.; Fusil,~S.; Briddon,~P.; Guiblin,~N.; others
  Electric-field control of magnetic order above room temperature. \emph{Nature
  materials} \textbf{2014}, \emph{13}, 345--351\relax
\mciteBstWouldAddEndPuncttrue
\mciteSetBstMidEndSepPunct{\mcitedefaultmidpunct}
{\mcitedefaultendpunct}{\mcitedefaultseppunct}\relax
\EndOfBibitem
\bibitem[Moty{\v{c}}kov{\'a} \latin{et~al.}(2023)Moty{\v{c}}kov{\'a}, Arregi,
  Sta{\v{n}}o, Pr{\r{u}}{\v{s}}a, {\v{C}}{\'a}stkov{\'a}, and
  Uhl{\'\i}{\v{r}}]{moty2023}
Moty{\v{c}}kov{\'a},~L.; Arregi,~J.~A.; Sta{\v{n}}o,~M.; Pr{\r{u}}{\v{s}}a,~S.;
  {\v{C}}{\'a}stkov{\'a},~K.; Uhl{\'\i}{\v{r}},~V. Preserving Metamagnetism in
  Self-Assembled FeRh Nanomagnets. \emph{ACS Applied Materials \& Interfaces}
  \textbf{2023}, \emph{15}, 8653--8665\relax
\mciteBstWouldAddEndPuncttrue
\mciteSetBstMidEndSepPunct{\mcitedefaultmidpunct}
{\mcitedefaultendpunct}{\mcitedefaultseppunct}\relax
\EndOfBibitem
\bibitem[Mattern \latin{et~al.}(2023)Mattern, von Reppert, Zeuschner, Herzog,
  Pudell, and Bargheer]{matt2023a}
Mattern,~M.; von Reppert,~A.; Zeuschner,~S.~P.; Herzog,~M.; Pudell,~J.-E.;
  Bargheer,~M. Concepts and use cases for picosecond ultrasonics with x-rays.
  \emph{Photoacoustics} \textbf{2023}, 100503\relax
\mciteBstWouldAddEndPuncttrue
\mciteSetBstMidEndSepPunct{\mcitedefaultmidpunct}
{\mcitedefaultendpunct}{\mcitedefaultseppunct}\relax
\EndOfBibitem
\bibitem[von Reppert \latin{et~al.}(2018)von Reppert, Willig, Pudell,
  R{\"o}ssle, Leitenberger, Herzog, Ganss, Hellwig, and Bargheer]{repp2018}
von Reppert,~A.; Willig,~L.; Pudell,~J.-E.; R{\"o}ssle,~M.; Leitenberger,~W.;
  Herzog,~M.; Ganss,~F.; Hellwig,~O.; Bargheer,~M. Ultrafast laser generated
  strain in granular and continuous FePt thin films. \emph{Applied Physics
  Letters} \textbf{2018}, \emph{113}, 123101\relax
\mciteBstWouldAddEndPuncttrue
\mciteSetBstMidEndSepPunct{\mcitedefaultmidpunct}
{\mcitedefaultendpunct}{\mcitedefaultseppunct}\relax
\EndOfBibitem
\bibitem[von Reppert \latin{et~al.}(2020)von Reppert, Willig, Pudell,
  Zeuschner, Sellge, Ganss, Hellwig, Arregi, Uhl{\'\i}{\v{r}}, Crut,
  \latin{et~al.} others]{repp2020}
von Reppert,~A.; Willig,~L.; Pudell,~J.-E.; Zeuschner,~S.; Sellge,~G.;
  Ganss,~F.; Hellwig,~O.; Arregi,~J.; Uhl{\'\i}{\v{r}},~V.; Crut,~A.; others
  Spin stress contribution to the lattice dynamics of FePt. \emph{Science
  advances} \textbf{2020}, \emph{6}, eaba1142\relax
\mciteBstWouldAddEndPuncttrue
\mciteSetBstMidEndSepPunct{\mcitedefaultmidpunct}
{\mcitedefaultendpunct}{\mcitedefaultseppunct}\relax
\EndOfBibitem
\bibitem[Reid \latin{et~al.}(2018)Reid, Shen, Maldonado, Chase, Jal, Granitzka,
  Carva, Li, Li, Wu, \latin{et~al.} others]{reid2018}
Reid,~A.; Shen,~X.; Maldonado,~P.; Chase,~T.; Jal,~E.; Granitzka,~P.;
  Carva,~K.; Li,~R.; Li,~J.; Wu,~L.; others Beyond a phenomenological
  description of magnetostriction. \emph{Nature communications} \textbf{2018},
  \emph{9}, 388\relax
\mciteBstWouldAddEndPuncttrue
\mciteSetBstMidEndSepPunct{\mcitedefaultmidpunct}
{\mcitedefaultendpunct}{\mcitedefaultseppunct}\relax
\EndOfBibitem
\bibitem[Chang \latin{et~al.}(2015)Chang, Wen, Chakraborty, Su, Zhang, Shuang,
  Nordlander, Sader, Halas, and Link]{chan2015}
Chang,~W.-S.; Wen,~F.; Chakraborty,~D.; Su,~M.-N.; Zhang,~Y.; Shuang,~B.;
  Nordlander,~P.; Sader,~J.~E.; Halas,~N.~J.; Link,~S. Tuning the acoustic
  frequency of a gold nanodisk through its adhesion layer. \emph{Nature
  Communications} \textbf{2015}, \emph{6}, 7022\relax
\mciteBstWouldAddEndPuncttrue
\mciteSetBstMidEndSepPunct{\mcitedefaultmidpunct}
{\mcitedefaultendpunct}{\mcitedefaultseppunct}\relax
\EndOfBibitem
\bibitem[Ju \latin{et~al.}(2004)Ju, Hohlfeld, Bergman, van~de Veerdonk,
  Mryasov, Kim, Wu, Weller, and Koopmans]{ju2004}
Ju,~G.; Hohlfeld,~J.; Bergman,~B.; van~de Veerdonk,~R.~J.; Mryasov,~O.~N.;
  Kim,~J.-Y.; Wu,~X.; Weller,~D.; Koopmans,~B. Ultrafast generation of
  ferromagnetic order via a laser-induced phase transformation in FeRh thin
  films. \emph{Physical review letters} \textbf{2004}, \emph{93}, 197403\relax
\mciteBstWouldAddEndPuncttrue
\mciteSetBstMidEndSepPunct{\mcitedefaultmidpunct}
{\mcitedefaultendpunct}{\mcitedefaultseppunct}\relax
\EndOfBibitem
\bibitem[Bergman \latin{et~al.}(2006)Bergman, Ju, Hohlfeld, van~de Veerdonk,
  Kim, Wu, Weller, and Koopmans]{berg2006}
Bergman,~B.; Ju,~G.; Hohlfeld,~J.; van~de Veerdonk,~R.~J.; Kim,~J.-Y.; Wu,~X.;
  Weller,~D.; Koopmans,~B. Identifying growth mechanisms for laser-induced
  magnetization in FeRh. \emph{Physical Review B} \textbf{2006}, \emph{73},
  060407\relax
\mciteBstWouldAddEndPuncttrue
\mciteSetBstMidEndSepPunct{\mcitedefaultmidpunct}
{\mcitedefaultendpunct}{\mcitedefaultseppunct}\relax
\EndOfBibitem
\bibitem[Radu \latin{et~al.}(2010)Radu, Stamm, Pontius, Kachel, Ramm, Thiele,
  D{\"u}rr, and Back]{radu2010}
Radu,~I.; Stamm,~C.; Pontius,~N.; Kachel,~T.; Ramm,~P.; Thiele,~J.-U.;
  D{\"u}rr,~H.; Back,~C. Laser-induced generation and quenching of
  magnetization on FeRh studied with time-resolved x-ray magnetic circular
  dichroism. \emph{Physical Review B} \textbf{2010}, \emph{81}, 104415\relax
\mciteBstWouldAddEndPuncttrue
\mciteSetBstMidEndSepPunct{\mcitedefaultmidpunct}
{\mcitedefaultendpunct}{\mcitedefaultseppunct}\relax
\EndOfBibitem
\bibitem[Li \latin{et~al.}(2022)Li, Medapalli, Mentink, Mikhaylovskiy, Blank,
  Patel, Zvezdin, Rasing, Fullerton, and Kimel]{li2022}
Li,~G.; Medapalli,~R.; Mentink,~J.; Mikhaylovskiy,~R.; Blank,~T.; Patel,~S.;
  Zvezdin,~A.; Rasing,~T.; Fullerton,~E.; Kimel,~A. Ultrafast kinetics of the
  antiferromagnetic-ferromagnetic phase transition in FeRh. \emph{Nature
  Communications} \textbf{2022}, \emph{13}, 2998\relax
\mciteBstWouldAddEndPuncttrue
\mciteSetBstMidEndSepPunct{\mcitedefaultmidpunct}
{\mcitedefaultendpunct}{\mcitedefaultseppunct}\relax
\EndOfBibitem
\bibitem[Pressacco \latin{et~al.}(2021)Pressacco, Sangalli, Uhl{\'\i}{\v{r}},
  Kutnyakhov, Arregi, Agustsson, Brenner, Redlin, Heber, Vasilyev,
  \latin{et~al.} others]{pres2021}
Pressacco,~F.; Sangalli,~D.; Uhl{\'\i}{\v{r}},~V.; Kutnyakhov,~D.;
  Arregi,~J.~A.; Agustsson,~S.~Y.; Brenner,~G.; Redlin,~H.; Heber,~M.;
  Vasilyev,~D.; others Subpicosecond metamagnetic phase transition in FeRh
  driven by non-equilibrium electron dynamics. \emph{Nature Communications}
  \textbf{2021}, \emph{12}, 5088\relax
\mciteBstWouldAddEndPuncttrue
\mciteSetBstMidEndSepPunct{\mcitedefaultmidpunct}
{\mcitedefaultendpunct}{\mcitedefaultseppunct}\relax
\EndOfBibitem
\bibitem[Kang \latin{et~al.}(2023)Kang, Omura, Yesudas, Lee, Lee, Lee,
  Taniyama, and Choi]{kang2023}
Kang,~K.; Omura,~H.; Yesudas,~D.; Lee,~O.; Lee,~K.-J.; Lee,~H.-W.;
  Taniyama,~T.; Choi,~G.-M. Spin current driven by ultrafast magnetization of
  FeRh. \emph{Nature Communications} \textbf{2023}, \emph{14}, 3619\relax
\mciteBstWouldAddEndPuncttrue
\mciteSetBstMidEndSepPunct{\mcitedefaultmidpunct}
{\mcitedefaultendpunct}{\mcitedefaultseppunct}\relax
\EndOfBibitem
\bibitem[Mariager \latin{et~al.}(2012)Mariager, Pressacco, Ingold, Caviezel,
  M{\"o}hr-Vorobeva, Beaud, Johnson, Milne, Mancini, Moyerman, \latin{et~al.}
  others]{mari2012}
Mariager,~S.~O.; Pressacco,~F.; Ingold,~G.; Caviezel,~A.;
  M{\"o}hr-Vorobeva,~E.; Beaud,~P.; Johnson,~S.; Milne,~C.; Mancini,~E.;
  Moyerman,~S.; others Structural and magnetic dynamics of a laser induced
  phase transition in FeRh. \emph{Physical Review Letters} \textbf{2012},
  \emph{108}, 087201\relax
\mciteBstWouldAddEndPuncttrue
\mciteSetBstMidEndSepPunct{\mcitedefaultmidpunct}
{\mcitedefaultendpunct}{\mcitedefaultseppunct}\relax
\EndOfBibitem
\bibitem[Quirin \latin{et~al.}(2012)Quirin, Vattilana, Shymanovich, El-Kamhawy,
  Tarasevitch, Hohlfeld, von~der Linde, and Sokolowski-Tinten]{qui2012}
Quirin,~F.; Vattilana,~M.; Shymanovich,~U.; El-Kamhawy,~A.-E.; Tarasevitch,~A.;
  Hohlfeld,~J.; von~der Linde,~D.; Sokolowski-Tinten,~K. Structural dynamics in
  FeRh during a laser-induced metamagnetic phase transition. \emph{Physical
  Review B} \textbf{2012}, \emph{85}, 020103\relax
\mciteBstWouldAddEndPuncttrue
\mciteSetBstMidEndSepPunct{\mcitedefaultmidpunct}
{\mcitedefaultendpunct}{\mcitedefaultseppunct}\relax
\EndOfBibitem
\bibitem[Mattern \latin{et~al.}(2023)Mattern, Jarecki, Arregi,
  Uhl{\'\i}{\v{r}}, R{\"o}ssle, and Bargheer]{matt2023b}
Mattern,~M.; Jarecki,~J.; Arregi,~J.~A.; Uhl{\'\i}{\v{r}},~V.; R{\"o}ssle,~M.;
  Bargheer,~M. Disentangling nucleation and domain growth during a
  laser-induced phase transition. \emph{arXiv} \textbf{2023}, \relax
\mciteBstWouldAddEndPunctfalse
\mciteSetBstMidEndSepPunct{\mcitedefaultmidpunct}
{}{\mcitedefaultseppunct}\relax
\EndOfBibitem
\bibitem[Willig \latin{et~al.}(2019)Willig, von Reppert, Deb, Ganss, Hellwig,
  and Bargheer]{will2019}
Willig,~L.; von Reppert,~A.; Deb,~M.; Ganss,~F.; Hellwig,~O.; Bargheer,~M.
  Finite-size effects in ultrafast remagnetization dynamics of FePt.
  \emph{Physical Review B} \textbf{2019}, \emph{100}, 224408\relax
\mciteBstWouldAddEndPuncttrue
\mciteSetBstMidEndSepPunct{\mcitedefaultmidpunct}
{\mcitedefaultendpunct}{\mcitedefaultseppunct}\relax
\EndOfBibitem
\bibitem[R{\"o}ssle \latin{et~al.}(2021)R{\"o}ssle, Leitenberger, Reinhardt,
  Ko{\c{c}}, Pudell, Kwamen, and Bargheer]{roes2021}
R{\"o}ssle,~M.; Leitenberger,~W.; Reinhardt,~M.; Ko{\c{c}},~A.; Pudell,~J.;
  Kwamen,~C.; Bargheer,~M. The time-resolved hard X-ray diffraction endstation
  KMC-3 XPP at BESSY II. \emph{Journal of Synchrotron Radiation} \textbf{2021},
  \emph{28}, 948--960\relax
\mciteBstWouldAddEndPuncttrue
\mciteSetBstMidEndSepPunct{\mcitedefaultmidpunct}
{\mcitedefaultendpunct}{\mcitedefaultseppunct}\relax
\EndOfBibitem
\bibitem[Pressacco \latin{et~al.}(2016)Pressacco, Uhl\'{\i}{\v{r}}, Gatti,
  Bendounan, Fullerton, and Sirotti]{pres2016}
Pressacco,~F.; Uhl\'{\i}{\v{r}},~V.; Gatti,~M.; Bendounan,~A.;
  Fullerton,~E.~E.; Sirotti,~F. Stable room-temperature ferromagnetic phase at
  the FeRh (100) surface. \emph{Scientific reports} \textbf{2016}, \emph{6},
  22383\relax
\mciteBstWouldAddEndPuncttrue
\mciteSetBstMidEndSepPunct{\mcitedefaultmidpunct}
{\mcitedefaultendpunct}{\mcitedefaultseppunct}\relax
\EndOfBibitem
\bibitem[Fan \latin{et~al.}(2010)Fan, Kinane, Charlton, Dorner, Ali, De~Vries,
  Brydson, Marrows, Hickey, Arena, \latin{et~al.} others]{fan2010}
Fan,~R.; Kinane,~C.~J.; Charlton,~T.; Dorner,~R.; Ali,~M.; De~Vries,~M.;
  Brydson,~R.~M.; Marrows,~C.~H.; Hickey,~B.~J.; Arena,~D.~A.; others
  Ferromagnetism at the interfaces of antiferromagnetic FeRh epilayers.
  \emph{Physical Review B} \textbf{2010}, \emph{82}, 184418\relax
\mciteBstWouldAddEndPuncttrue
\mciteSetBstMidEndSepPunct{\mcitedefaultmidpunct}
{\mcitedefaultendpunct}{\mcitedefaultseppunct}\relax
\EndOfBibitem
\bibitem[Zsoldos(1967)]{zsol1967}
Zsoldos,~L. Lattice Parameter Change of FeRh Alloys due to
  Antiferromagnetic-Ferromagnetic Transformation. \emph{physica status solidi
  (b)} \textbf{1967}, \emph{20}, K25--K28\relax
\mciteBstWouldAddEndPuncttrue
\mciteSetBstMidEndSepPunct{\mcitedefaultmidpunct}
{\mcitedefaultendpunct}{\mcitedefaultseppunct}\relax
\EndOfBibitem
\bibitem[Laskin \latin{et~al.}(2019)Laskin, Wang, Boschker, Braun, Srot, van
  Aken, and Mannhart]{lask2019}
Laskin,~G.; Wang,~H.; Boschker,~H.; Braun,~W.; Srot,~V.; van Aken,~P.~A.;
  Mannhart,~J. Magnetic properties of epitaxially grown SrRuO3 nanodots.
  \emph{Nano letters} \textbf{2019}, \emph{19}, 1131--1135\relax
\mciteBstWouldAddEndPuncttrue
\mciteSetBstMidEndSepPunct{\mcitedefaultmidpunct}
{\mcitedefaultendpunct}{\mcitedefaultseppunct}\relax
\EndOfBibitem
\bibitem[Rasche(2018)]{rasc2018}
Rasche,~C. Rapid contour detection for image classification. \emph{IET Image
  Processing} \textbf{2018}, \emph{12}, 532--538\relax
\mciteBstWouldAddEndPuncttrue
\mciteSetBstMidEndSepPunct{\mcitedefaultmidpunct}
{\mcitedefaultendpunct}{\mcitedefaultseppunct}\relax
\EndOfBibitem
\bibitem[Stephens and Malitson(1952)Stephens, and Malitson]{step1952}
Stephens,~R.~E.; Malitson,~I.~H. Index of refraction of magnesium oxide.
  \emph{Journal of Research of the National Bureau of Standards} \textbf{1952},
  \emph{49}, 249--252\relax
\mciteBstWouldAddEndPuncttrue
\mciteSetBstMidEndSepPunct{\mcitedefaultmidpunct}
{\mcitedefaultendpunct}{\mcitedefaultseppunct}\relax
\EndOfBibitem
\bibitem[Garcia-Vidal \latin{et~al.}(2022)Garcia-Vidal,
  Fern{\'a}ndez-Dom{\'\i}nguez, Martin-Moreno, Zhang, Tang, Peng, and
  Cui]{garc2022}
Garcia-Vidal,~F.~J.; Fern{\'a}ndez-Dom{\'\i}nguez,~A.~I.; Martin-Moreno,~L.;
  Zhang,~H.~C.; Tang,~W.; Peng,~R.; Cui,~T.~J. Spoof surface plasmon photonics.
  \emph{Reviews of Modern Physics} \textbf{2022}, \emph{94}, 025004\relax
\mciteBstWouldAddEndPuncttrue
\mciteSetBstMidEndSepPunct{\mcitedefaultmidpunct}
{\mcitedefaultendpunct}{\mcitedefaultseppunct}\relax
\EndOfBibitem
\bibitem[Kreibig and Vollmer(2013)Kreibig, and Vollmer]{kreibig2013}
Kreibig,~U.; Vollmer,~M. \emph{Optical properties of metal clusters}; Springer
  Science \& Business Media, 2013; Vol.~25\relax
\mciteBstWouldAddEndPuncttrue
\mciteSetBstMidEndSepPunct{\mcitedefaultmidpunct}
{\mcitedefaultendpunct}{\mcitedefaultseppunct}\relax
\EndOfBibitem
\bibitem[Maier \latin{et~al.}(2007)Maier, \latin{et~al.} others]{maier2007}
Maier,~S.~A.; others \emph{Plasmonics: fundamentals and applications};
  Springer, 2007; Vol.~1\relax
\mciteBstWouldAddEndPuncttrue
\mciteSetBstMidEndSepPunct{\mcitedefaultmidpunct}
{\mcitedefaultendpunct}{\mcitedefaultseppunct}\relax
\EndOfBibitem
\bibitem[Zoric \latin{et~al.}(2011)Zoric, Zach, Kasemo, and
  Langhammer]{zoric2011}
Zoric,~I.; Zach,~M.; Kasemo,~B.; Langhammer,~C. Gold, platinum, and aluminum
  nanodisk plasmons: material independence, subradiance, and damping
  mechanisms. \emph{ACS nano} \textbf{2011}, \emph{5}, 2535--2546\relax
\mciteBstWouldAddEndPuncttrue
\mciteSetBstMidEndSepPunct{\mcitedefaultmidpunct}
{\mcitedefaultendpunct}{\mcitedefaultseppunct}\relax
\EndOfBibitem
\bibitem[Kelly \latin{et~al.}(2003)Kelly, Coronado, Zhao, and
  Schatz]{kelly2003}
Kelly,~K.~L.; Coronado,~E.; Zhao,~L.~L.; Schatz,~G.~C. The optical properties
  of metal nanoparticles: the influence of size, shape, and dielectric
  environment. 2003\relax
\mciteBstWouldAddEndPuncttrue
\mciteSetBstMidEndSepPunct{\mcitedefaultmidpunct}
{\mcitedefaultendpunct}{\mcitedefaultseppunct}\relax
\EndOfBibitem
\bibitem[Chen and Lynch(1988)Chen, and Lynch]{chen1988}
Chen,~L.-Y.; Lynch,~D.~W. Ellipsometric studies of magnetic phase transitions
  of Fe-Rh alloys. \emph{Physical Review B} \textbf{1988}, \emph{37},
  10503\relax
\mciteBstWouldAddEndPuncttrue
\mciteSetBstMidEndSepPunct{\mcitedefaultmidpunct}
{\mcitedefaultendpunct}{\mcitedefaultseppunct}\relax
\EndOfBibitem
\bibitem[Rhee and Lynch(1995)Rhee, and Lynch]{rhee1995}
Rhee,~J.~Y.; Lynch,~D.~W. Optical properties of Fe-Rh alloys. \emph{Physical
  Review B} \textbf{1995}, \emph{51}, 1926\relax
\mciteBstWouldAddEndPuncttrue
\mciteSetBstMidEndSepPunct{\mcitedefaultmidpunct}
{\mcitedefaultendpunct}{\mcitedefaultseppunct}\relax
\EndOfBibitem
\bibitem[Saidl \latin{et~al.}(2016)Saidl, Brajer, Horák, Reichlová,
  Výborný, Veis, Janda, Trojánek, Maryško, Fina, Marti, Jungwirth, and
  Němec]{saidl2016}
Saidl,~V.; Brajer,~M.; Horák,~L.; Reichlová,~H.; Výborný,~K.; Veis,~M.;
  Janda,~T.; Trojánek,~F.; Maryško,~M.; Fina,~I.; Marti,~X.; Jungwirth,~T.;
  Němec,~P. Investigation of magneto-structural phase transition in FeRh by
  reflectivity and transmittance measurements in visible and near-infrared
  spectral region. \emph{New Journal of Physics} \textbf{2016}, \emph{18},
  083017\relax
\mciteBstWouldAddEndPuncttrue
\mciteSetBstMidEndSepPunct{\mcitedefaultmidpunct}
{\mcitedefaultendpunct}{\mcitedefaultseppunct}\relax
\EndOfBibitem
\bibitem[Ibarra and Algarabel(1994)Ibarra, and Algarabel]{ibar1994}
Ibarra,~M.; Algarabel,~P. Giant volume magnetostriction in the FeRh alloy.
  \emph{Physical Review B} \textbf{1994}, \emph{50}, 4196\relax
\mciteBstWouldAddEndPuncttrue
\mciteSetBstMidEndSepPunct{\mcitedefaultmidpunct}
{\mcitedefaultendpunct}{\mcitedefaultseppunct}\relax
\EndOfBibitem
\end{mcitethebibliography}

\end{document}